\def\aj{AJ}%
\def\araa{ARA\&A}%
\def\apj{ApJ}%
\def\apjl{ApJ}%
\def\apjs{ApJS}%
\def\aap{A\&A}%
\def\mnras{MNRAS}%
\def\jgr{J.~Geophys.~Res.}%
\def\planss{Planet.~Space~Sci.}%
\documentclass[a4paper,12pt]{article} 
\usepackage[margin=30mm]{geometry}
\usepackage[english]{babel}
\usepackage[pdftex]{graphicx} 
\usepackage[bf,small,nooneline]{caption}
\usepackage{citesupernumber} 
\usepackage{color}
\usepackage{url}
\usepackage{longtable}
\usepackage{pdflscape}
\usepackage{xcolor}

\usepackage{soul}     

\author{K.~G.\ Kislyakova\footnote{
  University of Vienna, Department of Astrophysics, Vienna, Austria; E-mail: kristina.kislyakova@univie.ac.at}, 
M.\ G\"{u}del\footnotemark[1], 
D.\ Koutroumpa\footnote{LATMOS-IPSL, CNRS, UVSQ Paris-Saclay, Sorbonne Universit\'{e}, Guyancourt, France}, \\
J.~A.\ Carter\footnote{Department of Physics and Astronomy, University of Leicester, Leicester, UK}, 
C.~M.\ Lisse\footnote{Johns Hopkins University Applied Physics Laboratory, Laurel, MD 20723, USA}, 
S.\ Boro~Saikia\footnotemark[1]
}
\title{\textbf{\textsf{
X-ray detection of astrospheres around three main-sequence stars and their mass-loss rates
}}}
\date{\today}

\begin{document}
\maketitle

\textit{Published in Nature Astronomy, DOI: 10.1038/s41550-024-02222-x}

\textbf{
Stellar winds of cool main sequence stars are very difficult to constrain observationally. One way to measure stellar mass loss rates is to detect soft X-ray emission from stellar astrospheres produced by charge exchange between heavy ions of the stellar wind and cold neutrals of the interstellar medium (ISM) surrounding the stars. Here we report detections of charge-exchange induced X-ray emission from the extended astrospheres of three main sequence stars, 70~Ophiuchi, $\epsilon$~Eridani, and 61~Cygni based on analysis of observations by XMM-Newton. We estimate the corresponding mass loss rates to be 66.5$\pm$11.1, 15.6$\pm$4.4, and 9.6$\pm$4.1 times the solar mass loss rate for 70~Ophiuchi, $\epsilon$~Eridani, and 61~Cygni, respectively, and compare our results to the hydrogen wall method. We also place upper limits on the mass loss rates of several other main sequence stars. This method has potential utility for determining the mass loss rates from X-ray observations showing spatial extension beyond a coronal point source. 
}



\section{Introduction}
\label{sec:intro}



Stellar winds play a crucial role for both stellar and planetary evolution. They carry away angular momentum and decelerate stellar rotation\cite{Weber67, Skumanich72,Wright11,Johnstone15}. In turn, the latter determines the general level of stellar activity and in particular the level of X-ray and extreme ultraviolet (XUV) radiation\cite{Pizzolato03}, which is the main driver of thermal escape of planetary atmospheres and therefore a crucial factor for planetary habitability\cite{Tu15,Johnstone21}. In addition, stellar winds directly drive various non-thermal atmospheric escape processes\cite{Gronoff20}, thus influencing the long-term evolution of planetary atmospheres.

Stellar mass loss rates can vary by 8 orders of magnitude or more, being generally lowest in the main sequence (MS) phase, and increasing drastically to upwards of 10$^{-5}$ to 10$^{-6}$ M$_\odot$/yr in the rapidly evolving asymptotic giant branch (AGB) stars, greatly contributing to the total decrease in a star's mass over just a few Myr. While there have been detections of winds from massive hot stars, giant stars, and T~Tauri stars in radio emission\cite{Bieging82, Scuderi98, Drake86, Cohen86}, near infrared hydrogen lines\cite{Folha01,Weigelt16}, and in the UV\cite{Conti78}, the very tenuous ionized winds of late-type MS stars are notoriously difficult to constrain. For MS stars, radio observations of the ionized winds have so far provided only upper limits\cite{Fichtinger17}. Other, albeit indirect, methods include estimates of stellar wind parameters based on Lyman-$\alpha$ observations of transiting exoplanets\cite{K14a,VidottoBourrier17}, but they are only available for stars hosting close-in planets and heavily depend on the models used for the planets' atmospheres. 

For very young and rapidly rotating stars, it is possible to observe slingshot prominences visible as transient absorption features in the Balmer and Ca II H\& K lines\cite{Jardine19,Jardine20}. Mass loss rates estimated for these stars are usually several orders of magnitude higher than the solar mass loss rate of $2 \times 10^{-14}$~M$_\odot$/year\cite{Johnstone15}. However, such observations are possible only for a minority of MS stars. Detections of episodic Coronal Mass Ejections (CMEs) at 10$^3$ to 10$^4$ times quiescent rates have also proved to be difficult\cite{Odert17, Crosley18}. 
For CME detections, indirect methods have been developed based on the Doppler signal produced by CME-related plasma motions\cite{Argiroffi19} and on sudden dimmings in the extreme ultraviolet and X-ray emission\cite{Veronig21}. In the future, these methods might shed more light on the mass loss rates due to CMEs and the role CMEs play in the overall mass loss rate.

Another method to detect stellar winds is based on observations of charge exchange between stellar winds and the surrounding interstellar neutral medium. 
The protons of the fully ionized stellar wind can charge exchange (CX) when they collide with neutral atoms and molecules in the astrosphere. In the process, they take away an electron from the neutral, producing a slow ion and a fast neutral hydrogen atom called an Energetic Neutral Atom (ENA). Since ENAs keep the initial energies of their stellar wind ion precursor, they move with much higher speeds than the typical interplanetary hydrogen atom and and thus can be detected in the wings of the host star's Lyman-$\alpha$ emission line. The central part of the line is not observable from near-Earth orbit due to absorption by the interstellar medium (ISM) and contamination by the geocorona. The wings of the line, on the other hand, are available for analysis. So far, the most fruitful method of stellar wind detection was based on observations of the Ly$\alpha$ absorption by a neutral H wall formed around astrospheres in the ISM. This ``hydrogen wall method'' has provided us with the most currently available mass loss rate estimates\cite{Wood02,Wood05,Wood21}.

Stellar winds also contain a fraction of highly ionized bare, hydrogenic (one remaining electron) and heliogenic (two-remaining electron) heavy elements such as C, N, and O\cite{Schwadron00}. Just like protons, these ions can also charge exchange when they collide with a neutral. However, instead of becoming an ENA, they capture an electron into an excited state. The electron then falls onto the lowest available orbit producing an X-ray or extreme ultraviolet (EUV) photon. The EUV part of the CX spectrum is only observable for solar system objects given the ISM absorption, but soft X-rays from other stellar systems generated by bare and hydrogenic ions can be observed out to tens of parcecs. In the solar system, solar wind charge exchange (SWCX) emission has been observed from planets and comets\cite{Lisse96,Lisse01,Lisse04,Bhardwaj07} and provides a natural laboratory to study the solar wind's composition\cite{Schwadron00}. It is reasonable to assume that similar charge exchange processes should take place in exoplanetary systems. Although theoretical estimates indicate that soft X-rays from close-in exoplanets cannot be observed with current instruments\cite{K15}, stellar astrospheres present a much better opportunity for observations due to the very large CX interaction region volume. Here, the neutrals are particles intruding from the local, mostly neutral ISM into the solar wind. Previously, SWCX was proposed as a potentially powerful tool for detecting stellar winds\cite{Wargelin01}, and was used to put an upper limit on the wind of Proxima Centauri using \textit{Chandra} observations\cite{Wargelin02}. In this article, we present the first detections of SWCX emission from the astrospheres of three main sequence stars and estimate stellar mass loss rates based on these observations. 

We present our analysis of X-ray observations by the XMM-Newton observatory\cite{Jansen01} of the stars 70 Ophiuchi (70~Oph), $\epsilon$~Eridani ($\epsilon$~Eri), $\alpha$ Centauri ($\alpha$~Cen), Proxima Centauri (Prox Cen), Procyon, $\tau$~Ceti ($\tau$~Cet), and 61~Cygni (61~Cyg). For three of these stars, we detect excess X-ray emission from their astrospheres and interpret it as an astrospheric signal. Based on observed excess CX emission, we estimate corresponding mass loss rates of these stars. 

\section{Results}
\label{sec:results}

\tiny
\begin{longtable}{llllllllll}
\hline\hline
Star      & ST & D & $R_{\rm s}$  & $\dot{M}$   & L$_X^{star}$   & L$_X^{astro}$  & $\theta$ & $\beta$ & $\dot{M}_{\rm HW}$ \\ 
      &  & (pc) & ($R_\odot$) & ($\dot{M}_{\odot}$)  & (erg/s)  & (erg/s) & (deg) &  (deg) & ($\dot{M}_{\odot}$) \\ \hline

70~Oph          & K0~V+K5~V & 5.12 & 0.91(A)\cite{Bruntt10}+0.71(B)\cite{Fracassini01} & 66.5$\pm$11.1 & 1.77$\times 10^{28}$ & 6.4$\times 10^{26}$ & 120 & 4.2 & 100 \\
$\epsilon$~Eri (Full Frame)  & K2~V & 3.22 & 0.86\cite{Fracassini01} & 15.6$\pm$4.4  & 1.4$\times 10^{28}$ & 7.2$\times 10^{25}$ & 76 & 68.9 & 30 \\
$\epsilon$~Eri (Large Window)  & & &  &  17.9$\pm$10.6   & 1.2$\times 10^{28}$  & 4.1$\times 10^{25}$ &    &   &    \\               
61~Cyg          & K5~V+K7~V   & 3.49 & 0.665(A)+0.595(B)\cite{Kervella08} & 9.6$\pm$4.1 & 2.4$\times 10^{27}$  & 6.4$\times 10^{25}$ & 46 & NA & 0.5  \\
$\alpha$~Cen    & G2~V+K0~V  &  1.34 & 1.225(A)+0.864(B)\cite{Bruntt10} & $<$0.75 & 1.9$\times 10^{27}$ & -- & 79  & 158 & 2.0  \\       
Procyon         & F5~IV--V + DQZ  & 3.51 & 2.059\cite{Bruntt10} & $<$0.8 & 7.0$\times 10^{27}$ & -- & NA & NA & -- \\
Prox~Cen        & M5.5~V & 1.30 & 0.1542\cite{Kervella17} & $<$0.7 & 7.6$\times 10^{26}$ & -- & 79 & 161 & $<$0.2  \\
$\tau$~Cet    & G8~V  &  3.65 & 0.794\cite{Bruntt10} & $<$4.2 & 1.9$\times 10^{26}$ & -- & 122.9 & NA & -- \\ \hline

\caption{List of analyzed XMM-Newton observations, stellar parameters\cite{Wood02,Wood05} and distances (from the GAIA DR3 archive), estimated mass loss rates, and upper limits. The angle $\theta$ is the orientation angle, that is, the angle between the upwind direction of the ISM flow seen by the star and our line of sight\cite{Wood05}. The column $\dot{M}_{\rm HW}$ presents available MLRs based on the hydrogen wall method\cite{Wood05} (only A component for 61~Cyg). The angle $\beta$ is the position angle of the astrosphere, that is, the projection angle of the astrosphere on the XMM-Newton's FOV showing the tail directions in degrees east (left) in RA/DEC coordinates (Brian Wood, private communication; $\beta$ is not available for all stars). We assume the solar mass loss rate of $\dot M_\odot = 2 \times 10^{-14}~M_\odot~\rm{year}^{-1}$. The errors only reflect spectral fitting uncertainties; see discussion following Eqs.~\ref{e:wind} and \ref{e:mdot} regarding other systematic uncertainties. }
\label{t:results}
\end{longtable}
\normalsize

Detecting a signal from a stellar astrosphere is not a trivial task due to faintness of these objects and difficulties related to spatial resolution and background emission. For these reasons, we have limited our search to stars examined in previous Lyman-$\alpha$ studies\cite{Wood02,Wood05} that lie within approximately 5~pc from the Sun. Since stellar astrospheres are extended objects and cover a significant portion of the field-of-view (FOV) of current X-ray observatories for such close targets, we have selected archival XMM-Newton observations in Full Frame mode (with one additional observation in Large Window mode) to search for astrospheric signals. This has limited our search to 61 Cyg, 70~Oph, $\epsilon$~Eri, $\alpha$~Cen, Prox Cen, Procyon, and $\tau$~Cet. Table~\ref{t:results} summarizes the stellar properties of these systems. For binary stars, our analysis presents the mass loss rate estimates from both components together. For $\epsilon$~Eri, we show two separate estimates of mass loss rates based on two observations in different modes (Full Frame and Large Window). For other stars, the mass loss rates or upper limits are based on multiple observations (the average of all upper limits), except $\tau$~Cet where only one observation was available.

We have used the standard tools developed for the data reduction and calibration of XMM-Newton observations, namely, the XMM-Newton Science Analysis System (SAS). For science analysis, we applied the X-Ray Spectral Fitting Package (\texttt{Xspec}) created by the NASA's High Energy Astropyhsics Science Archive Research Center (see Code availability section for details).
 
\begin{figure*}[ht]
\includegraphics[width=1.0\columnwidth]{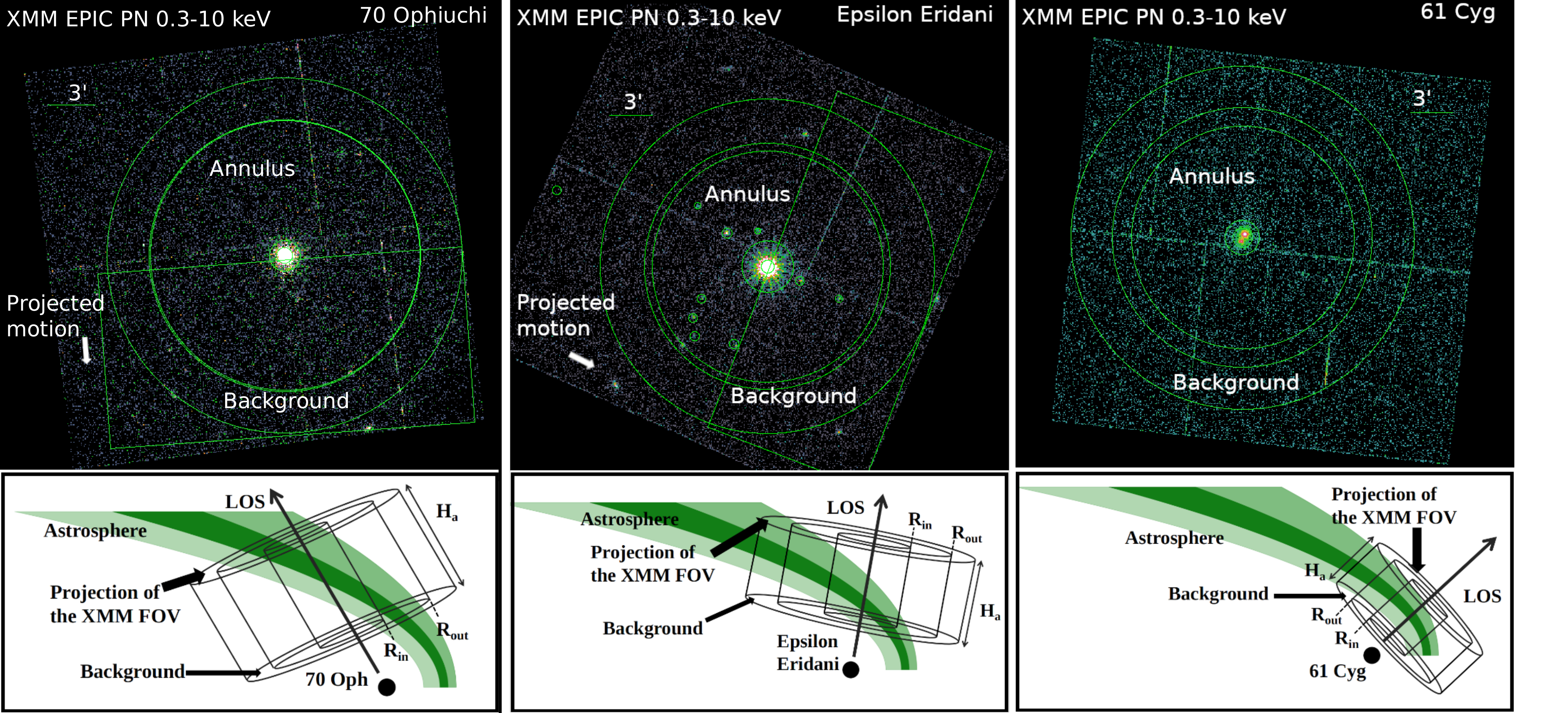}
\caption{Top: Images around the stars 70 Oph (observation 0044740101), $\epsilon$ Eri (observation 0112880501), and 61 Cyg (observation 0801871001), taken with the XMM-Newton EPIC PN camera. The central circle shows the regions used to extract the spectra of the central stars. The regions marked as ``Annulus'' show the areas where we have searched for the astrospheric signal. The part of the far annulus that is within the green box and is labeled ``background'' (approximately half of the annulus) was used for background subtraction. The white arrows point in the approximate direction of the projected stellar motion based on position angles of the astrospheres, 4.2$^\circ$ for 70~Oph, and 68.9$^\circ$ for $\epsilon$~Eri. For 61~Cyg, the position angle of the astrosphere is unknown. For this reason, we have used the full far annulus as a background. Bottom: observed line-of-sights for 70 Oph, $\epsilon$~Eri and 61~Cyg\cite{Wood02,Wood05} and projection of the annuli in the XMM FOV, with $R_{in}$, $R_{out}$ and $H_a$ being the CX region's inner and outer boundaries and height, respectively. The astrospheres and the stars are not plotted to scale. Different shades of green illustrate regions of different H densities in the astrospheres.  
}
\label{fig:StarsRegions}
\end{figure*}

We have developed a procedure to work with extended astrospheric sources that is described in more details in Methods. Here, we present a short summary. First, we have extracted the spectrum of the central star following the procedures for point-like sources described in the SAS threads for the three  European Photon Imaging Cameras (EPIC) of XMM-Newton, namely, PN, MOS1, and MOS2. We have also created images of the observations as described in the SAS threads. We have handled the images using the SAOImageDS9 software, as is the standard procedure. We then extracted the spectra from a wide annulus around each star as illustrated by Fig.~\ref{fig:StarsRegions}. We extracted the background from outside the outer radius of the annulus. Out-of-time events have been removed from the images and from the spectra.

\begin{figure*}[ht]
\includegraphics[width=1.0\columnwidth]{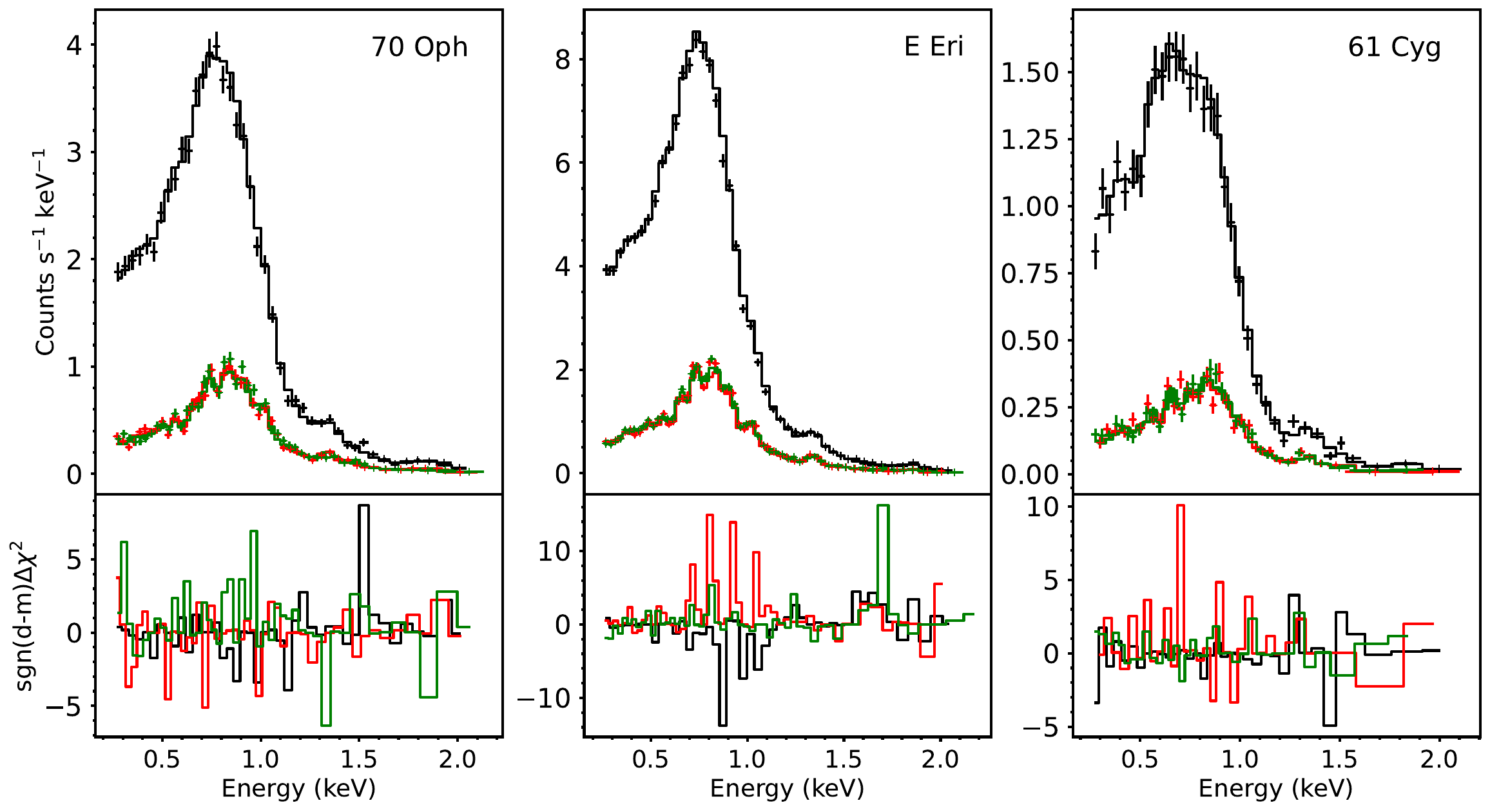}
\caption{Stellar spectra of 70~Oph, $\epsilon$~Eri, and 61~Cyg fitted with 3T-vapec models, with the $\chi^2$ shown in the lower panels. The data are presented as mean values $\pm 1.64 \sigma$ (90\% confidence interval). The number of bins for statistics was 161 (70 Oph), 169 ($\epsilon$~Eri), and 105 (61~Cyg). The spectra are shown for the same observations as in Fig.~\ref{fig:StarsRegions}. The black error bars and the black histogram show the spectra and the fit, respectively, obtained with the PN camera, while the green and red error bars and histograms show the spectra and model fits from the MOS1 and MOS2 cameras, respectively.
} \label{fig:StarsSpectra}
\end{figure*}

\begin{figure*}[ht]
\includegraphics[width=1.0\columnwidth]{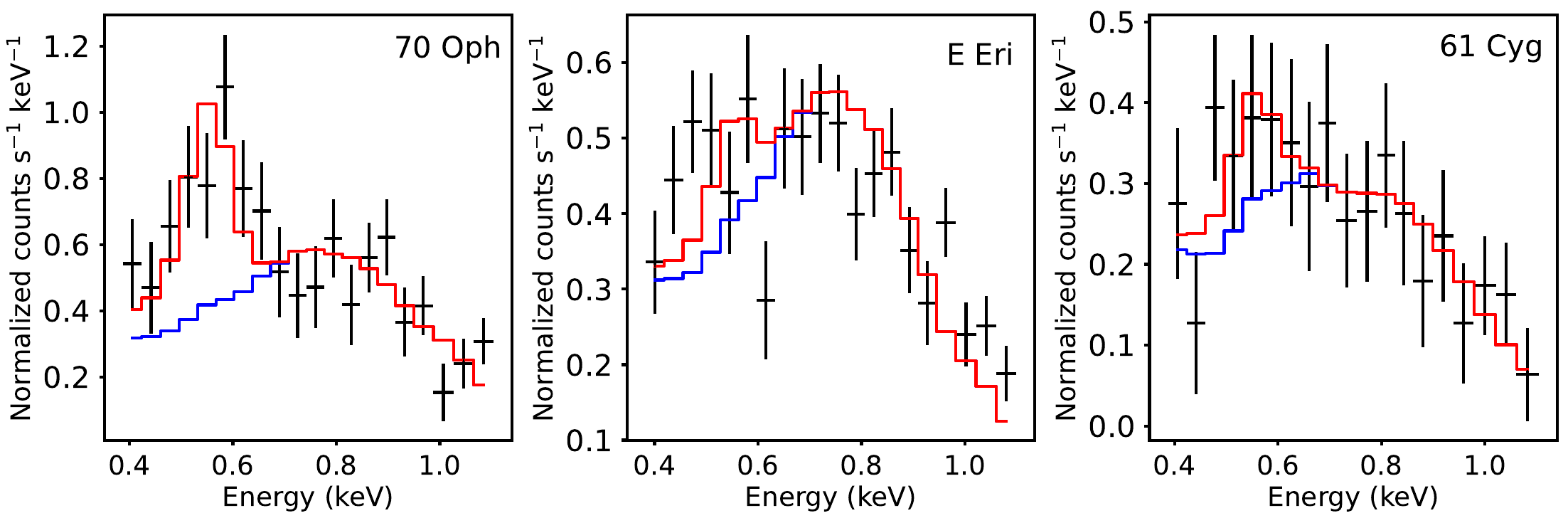}
\caption{Spectra of the annuli surrounding 70~Oph (observation 0044740101), $\epsilon$ Eri (observation 0112880501), and 61~Cyg (observation 0801871001). Only PN data are shown for clarity. The data are presented as mean values $\pm 1.64 \sigma$ (90\% confidence interval). The number of bins for statistics are shown in Table~\ref{t:annuli}. The black error bars shown the data. The blue solid lines show the scaled down stellar models from Fig.~\ref{fig:StarsSpectra}. The red solid lines shown the same model, but with an added Gaussian line at 0.56~keV. The excess signal resembles in shape the typical SWCX signal observed in the solar system\cite{Snowden04}. 
} \label{fig:Annuli}
\end{figure*}

\begin{figure*}[ht]
\includegraphics[width=1.0\columnwidth]{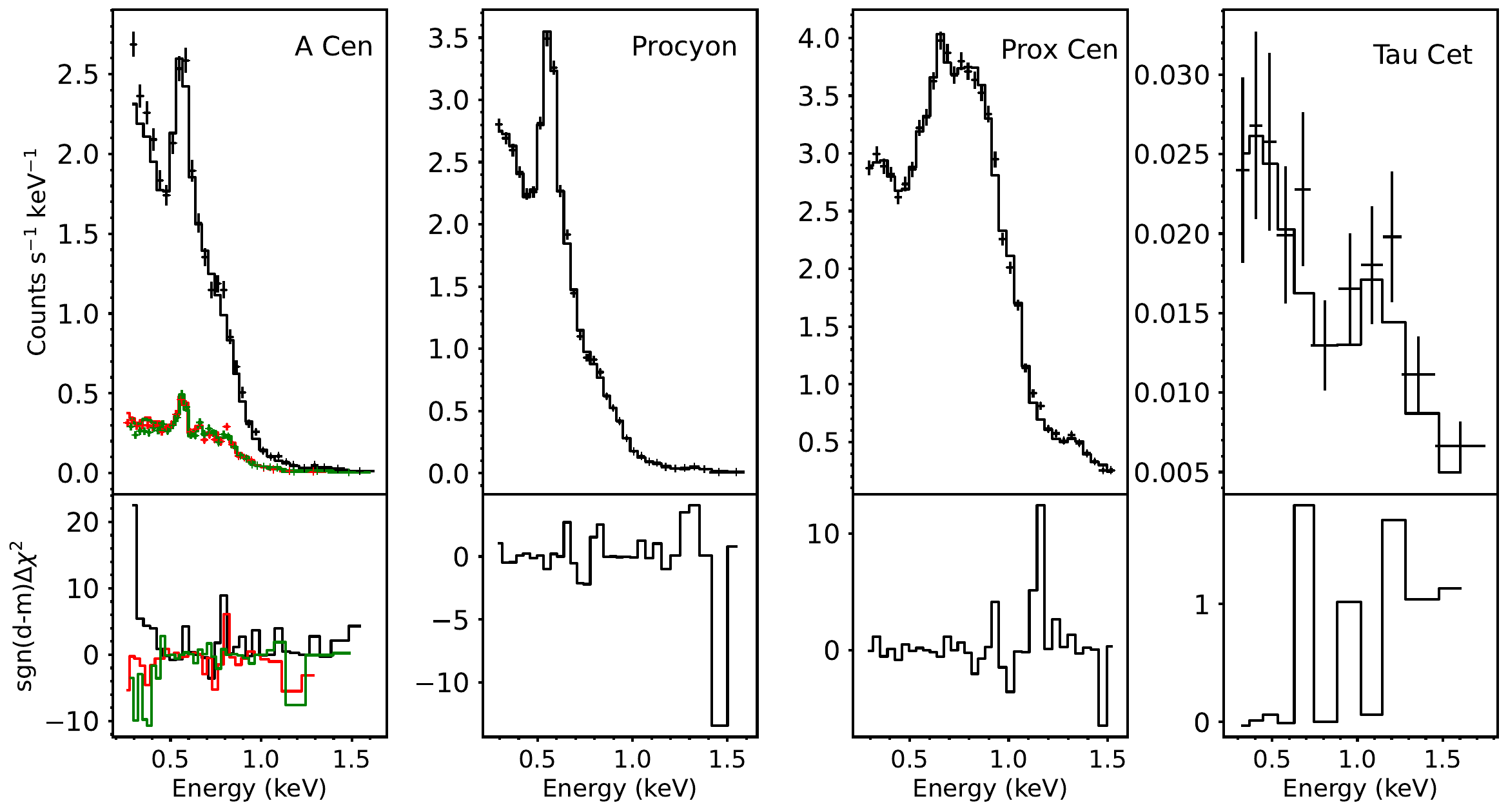}
\caption{Stellar spectra of Alpha Cen (obs ID 0760290301), Procyon (obs ID 0415580201), Prox Cen (obs ID 0801880501), and Tau Cet (obs ID 0670380501) fitted with 3T-vapec models, with the $\chi^2$ shown in the lower panels. The data are presented as mean values $\pm 1.64 \sigma$ (90\% confidence interval). The number of bins for statistics was 94 (Alpha Cen), 34 (Procyon), 63 (Prox Cen), and 18 (Tau Cet). The black error bars and the black histogram show the spectra and the fit, respectively, obtained with the PN camera, while the green and red error bars and histograms show the spectra and model fits from the MOS1 and MOS2 cameras, respectively.
} \label{fig:StarsSpectraNoDet}
\end{figure*}

\begin{figure*}[ht]
\includegraphics[width=1.0\columnwidth]{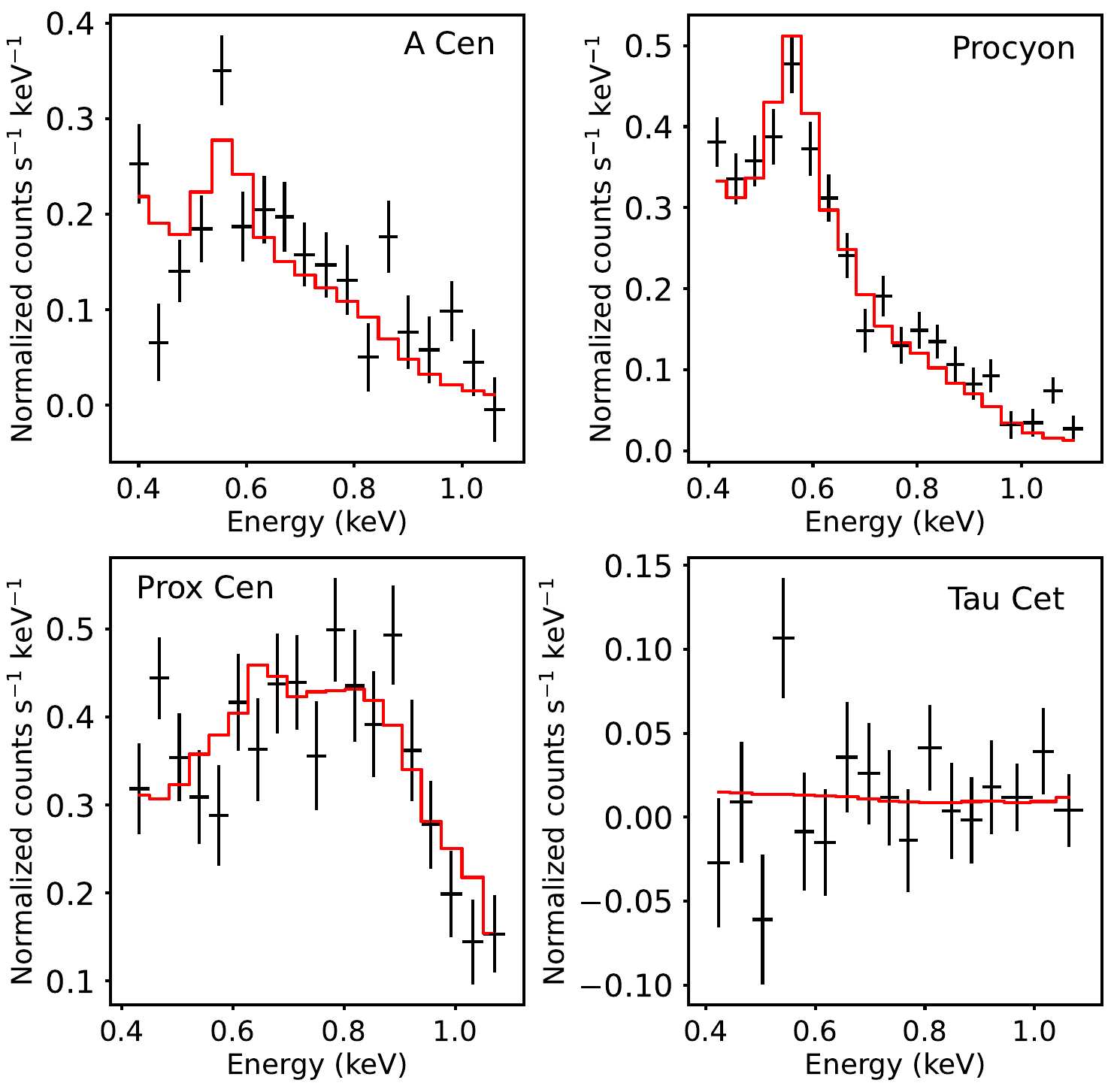}
\caption{Spectra of the annuli surrounding Alpha Cen, Procyon, Prox Cen, and Tau Cet, for the same observations shown in Fig.~\ref{fig:StarsSpectraNoDet}. 
The data are presented as mean values $\pm 1.64 \sigma$ (90\% confidence interval). The number of bins for statistics are shown in Table~\ref{t:annuli}. Only PN data are shown for clarity. The black error bars shown the data. The red solid lines shown the vapec+vapec+vapec model with an added Gaussian line at 0.56~keV. Only the red line is visible because the two models with and without the additional Gaussian line overlap thus indicating that no astrospheric CX signal has been detected. The models overlap because the best fit of the data is achieved for a Gaussian line with zero norm, which indicates that adding any additional flux around 0.56~keV in comparison to the one predicted by the stellar model only worsens the fit.}
\label{fig:AnnuliNoDet}
\end{figure*}

We have searched for residual excess X-ray emission in the annulus produced by charge exchange between stellar winds and the neutrals from the ISM as follows. Because the noise contribution to the stellar signal in the wings of the PSF is statistically independent of the noise in the central part of the stellar PSF, we fitted the stellar spectrum from the central PSF with a coronal model and also studied the quality of the fit to ascertain that there are no significant systematic fit residuals that may imitate an SWCX signal. We have then fitted the height of this best fit stellar spectrum to the annulus spectrum and subtracted it (see Methods). We looked for an excess emission in the 0.56~keV oxygen K$\alpha$ triplet line above the scaled-down stellar coronal flux. Fig.~\ref{fig:StarsSpectra} shows stellar spectra of 70~Oph, $\epsilon$~Eri, and 61~Cygi fitted with a coronal model comprising three temperature components (3T vapec+vapec+vapec model) in XSPEC. The vapec model represents an emission spectrum from collisionally-ionized gas calculated from the AtomDB atomic database (\url{http://atomdb.org/}). The 3T vapec model synthesizes a spectrum of a three-component plasma with three different temperatures. We have kept the element abundances between the components equal. Figs.~\ref{fig:Annuli} show the spectra of the annuli around the same stars as in Fig.~\ref{fig:StarsSpectra}. For the fits, all parameters of the model were kept fixed at the stellar values except the flux normalization of the model. This is important because most of the counts in the annuli are expected to be stellar X-rays scattered into the wings of the stellar point spread function. The spectral fits (blue histograms) confirm this, but for some stars, there is also some excess emission around the oxygen K$\alpha$ line that cannot be explained by the spectrum of the central star. We have focused on the oxygen K$\alpha$ line as it is one of the brightest CX lines from high abundance oxygen ions in solar and stellar winds\cite{Wargelin01}. The quality of the spectra did not allow us to search for multiple weaker lines of various elements. Stellar and annuli spectra for stars where we could not detect an astospheric signal are shown in Figs. \ref{fig:StarsSpectraNoDet}, \ref{fig:AnnuliNoDet}.

To search for the excess emission, we then added a Gaussian line at 0.56~keV to the model and fitted its flux. If this procedure improved the fit for PN, MOS1, and MOS2 spectra fitted together (where all three were available), and if the excess (red histograms) was significant, meaning that the estimated fluxes exceeded the errors and that adding the line improved the $\chi^2$ statistics, we interpreted this as a sign of astrospheric SWCX emission. We then proceeded to estimate the flux in this line alone without the contribution from the stellar PSF and interpreted this as the flux from the stellar astrosphere in the K$\alpha$ line. Fig.~\ref{fig:Annuli} shows the spectra of the annuli around 70~Oph, $\epsilon$~Eri, and 61~Cyg fitted with scaled down stellar models with and without the additional Gaussian line. For the three spectra shown in this figure, adding the line always improved the fit. 
If adding the additional Gaussian line did not improve the fit, that is, if the best with was achieved with for a zero norm of the line, or if the error on estimated SWCX flux exceeded the flux itself, we interpreted the flux in the added 0.56~keV line as a non-detection and estimated an upper limit for the mass loss rate based on the error of the flux.

Table~\ref{t:results} presents the summary of the stellar parameters, the measured coronal and astrospheric X-ray fluxes (this study), and our our estimated mass loss rates (MLRs) and upper limits. For stars where multiple observations were available, the mass loss rates and errors (or the upper limits) are the average values of all observations with equal weights. The sensitivity of our method depends on the distance to the star and the quality of the spectra. For this reason, the upper limit for Prox Cen is much lower than the upper limit for the more distant $\tau$~Cet. For binary stars with two MS components, namely, 70 Oph and 61 Cyg our analysis presents the combined mass loss rate estimates from both components. 

For 70~Oph, the fits were improved by adding the additional Gaussian line for all annuli in all three observations and for all three EPIC cameras fitted together and separately. For 61~Cyg and and $\epsilon$~Eri, adding the line always improved the fit for all three cameras fitted together. In some cases, the fit was not improved for MOS1 or MOS2 fitted separately, but we still interpret it as a detection when fitted jointly for all detectors (see Table~\ref{t:annuli} for $\chi^2$-statistic for all observations). To increase the robustness of our detection from for $\epsilon$~Eri which would otherwise be based on just one observation in Full Frame mode, we have also analyzed one additional observation in Large Window mode (listed in Table~\ref{t:results} separately). Only the PN camera was available for analysis in this observation. This observation also yielded a detection corresponding to a very similar mass loss rate. For other stars, we did not obtain any detections. 

Knowing the excess X-ray luminosity in the annulus $a$, $F_{Xa}$ in erg/s, we have estimated the flux of wind particles in this annulus, $F_{Wa}$ in particles/s/cm$^2$, based on the number of heavy oxygen ions that have undergone CX in this region:

\begin{equation}
F_{Wa} = F_{Xa} / (\sigma n_H  f_{O}  V_{CX}  E_{ph} ),
\label{e:wind}
\end{equation}
where $\sigma = 4 \times 10^{-15}$~cm$^2$ is the assumed velocity and species-independent CX cross section for oxygen ion charge exchange\cite{Koutroumpa07}, $n_H$ is the density of neutral hydrogen CX targets, $f_O$ is the sum of the fractions of O$^{8+}$ and O$^{7+}$ ions in the stellar wind. We have assumed a value equal to a slow solar wind state of $f_O = 1.272 \times 10^{-4} + 4.77 \times 10^{-5}$ for H-like and bare oxygen ions, respectively\cite{Wargelin01}. We note that the O$^{8+}$/O$^{7+}$ ratio can vary significantly depending on the charge state and temperature of the wind\cite{Kharchenko98,Schwadron00}. Furthermore, $V_{CX} = H_a (R^2_{out} - R^2_{in})$ is the volume of the cylindrical CX region in each annulus with $R_{in}$ and $R_{out}$ being the annulus' inner and outer boundaries, respectively; $H_a$ is the annulus height, and $E_{ph} = 8.97 \times 10^{-10}$~erg is the energy of one photon in the oxygen K$\alpha$ line. The mass loss rate in the annulus around the star can then be estimated as 
\begin{equation}
\dot M_{a} = 4 \pi F_{Wa} D_{a}^2 m_{w},
\label{e:mdot}
\end{equation}
where $m_{w}$ is an average mass of a wind particle assumed to equal 1.3 proton masses, and $D_{a}$ is the average distance from the star to the CX region which we assumed to be equal to the distance to the center of the annulus.  We assumed that each heavy ion produces one X-ray photon. The largest errors on the $\dot{M}$ estimate come from the uncertainties in $n_H$ and $V_{CX}$, scaling up to a factor of a few. We note that the spectral fitting uncertainties are the ones relevant to determining the significance of any SWCX emission excess, while the larger systematic uncertainties in Eqs.~\ref{e:wind} and \ref{e:mdot} pertain to translation of that excess into a mass loss rate. One of the difficulties of interpreting the observations comes from looking at 2D projections of stellar astrospheres that in reality are complex 3D objects (see lower panels of Fig.~\ref{fig:StarsRegions}). So, we detect excess CX emission produced in a cylinder around the star with a finite height along the LOS, which contains regions with different densities of both wind particles and neutral ISM. We have used a simplified estimate for the height of the CX cylinder in Eq.~\ref{e:mdot} equal to $H_a = 3 D_a$, where $D_{a}$ is the annulus' center $D_a = (R_{out} - R_{in})/2$. We have chosen $R_{in}$ and $R_{out}$ such that a significant portion of the cameras' FOV would be included. Furthermore, we assumed the density of neutral n$_H$ atoms of 0.5~cm$^{-3}$ in Eq.~\ref{e:wind}. The values of $H_a$ and $n_H$ were based on average values one can expect based on reconstructions of the stellar astrospheres of interest based on the hydrogen wall method\cite{Wood02,Wood05}. Such reconstructions were available for 70~Oph, $\alpha$~Cen, 61~Cyg, and $\epsilon$~Eri\cite{Wood02,Wood05}. In the future, a detailed reconstruction of stellar astrospheres including precise modeling of the $n_H$ distribution will be beneficial, but such modeling is beyond the scope of the current article.

\footnotesize
\begin{longtable}{lllllll}
\hline\hline
Star  &  Observation ID  & MLR & \begin{tabular}{@{}c@{}}$R_{in}$--$R_{out}$ (AU) \\Annulus \end{tabular}  & \begin{tabular}{@{}c@{}}$R_{in}$--$R_{out}$ (AU) \\Background \end{tabular} & $\chi^2$/bins & $\chi^2_{0.56~keV}$/bins  \\
\hline
70~Oph &  \textbf{0044740101} & 79.2$\pm$10.5 & 307--2687 & 2713--3546 & 50.83 / 21   &  19.46 / 21  \\
       &  \textbf{0044741301} & 63.8$\pm$12.5 & 307--2687 & 2713--3546 &  48.94 / 21  & 23.02 / 21   \\     
       &  \textbf{0044741401} & 56.6$\pm$10.4 & 307--2687 & 2713--3546 & 43.40 / 20  &  24.39 / 20  \\     \hline         

$\epsilon$~Eri  & \textbf{0112880501} & 15.6$\pm$4.4 & 321--1447  & 1544--2096  & 58.74 / 21  & 52.66 / 21 \\
                &  \textbf{0748010101} & 17.9$\pm$10.6 & 592--1125 & BOX &   17.98 / 16  &  15.09 / 16    \\      \hline

61~Cyg   & \textbf{0690800901}  & 4.6$\pm$4.2 & 244--1570 & 1832--2417 & 28.13 / 20  & 26.83 / 20  \\
         & \textbf{0741700601} & 7.5$\pm$3.4& 244--1570 & 1832--2417 & 21.86 / 20 &  19.71 / 20 \\
         & \textbf{0801871001} & 16.7$\pm$4.7 & 244--1570 & 1832--2417 &  12.89 / 20 &  10.23 / 20   \\     \hline

$\alpha$~Cen  & 0550060901 & $<$1.0 & 80--402 & 536--670  & 68.63 / 20  & 68.63 / 20  \\
              & 0550061001 & $<$0.15 & 80--402 & 536--670  & 42.97 / 19  & 42.97 / 19  \\
              & 0670320301 & $<$0.93 & 80--402 & 536--670  & 43.29 / 18  & 43.29 / 18  \\
              & 0690800301 & $<$0.66 & 80--402 & 536--670  & 17.21 / 21  & 17.21 / 21  \\              
              & 0741700301 & $<$1.4 & 80--402 & 536--670  &  41.12 / 21 & 41.12 / 21  \\
              & 0760290301 & $<$0.3 & 80--402 & 536--670  & 53.64 / 19  & 53.64 / 19 \\      \hline

              
Prox~Cen$^*$  & 0551120201 & $<$0.96 & 61--377 & 377--455 & 17.33 / 21 & 16.82 / 21  \\
              & 0551120301 & $<$0.51 & 61--377 & 377--455 &  15.94 / 21 & 15.94 / 21  \\       
              & 0801880501 & $<$0.64  & 61--377 & 377--455 &  25.66 / 21  & 25.66 / 21  \\       \hline
              
Procyon$^*$  & 0415580101 & $<$0.7 & 164--1018  & 1017--1228  & 41.96 / 21 & 41.96 / 21    \\
             & 0415580201 & $<$0.64 & 164--1018  & 1026--1228  & 39.40 / 20 & 39.40 / 20   \\
             & 0415580301 & $<$1.08 & 164--1018  & 1017--1228  & 19.86 / 21 & 19.86 / 21   \\ \hline

$\tau$~Cet$^*$ & 0670380501 & $<$4.2 & 128--1277 & 1368--1825  & 19.68 / 19 &  19.56 / 19  \\      \hline 

\caption{Sizes of CX regions, estimated MLRs, $\chi^2$ statistics and the number of bins for each observation for PN, MOS1, and MOS2 cameras fitted together. The inner ($R_{in}$) and outer ($R_{out}$) edges of each annulus are shown in astronomical units. The heights of the cylindrical CX regions, H$_a$, was assumed equal to three times the average radii of the annuli. For observations marked with an asterisk, we only relied on the PN camera to obtain estimates because astrospheric annuli were not sufficiently covered on MOS1 and MOS2 FOVs. The values for $\chi^2$ statistics were calculated for the 0.4--1.2 keV energy range. The table also shows the number of bins these $\chi^2$ values were calculated for. Observations that yielded detections are marked with bold font. The observation 0748010101 of $\epsilon$~Eri in Large Window had a non-standard background, namely, a large box far away from the central star.} 
\label{t:annuli}
\end{longtable}
\normalsize

In addition to the size of the annulus, the orientation of the astrosphere matters. It can be described by the two angles, $\theta$ and $\beta$. The former is the angle between the upwind direction of the ISM flow seen by the star and our line of sight, while the latter is the position angle of the astrosphere, that is, the projection angle of the astrosphere on the XMM-Newton’s FOV showing the tail directions in degrees (see Table~\ref{t:results}). The angles $\beta$ and $\theta$ are not available for all stars of our sample. For stars where the position angle $\beta$ was available, we have selected the background as a wide half-annulus beyond the outer radius of the astrospheric annulus in the FOV in the upwind direction of the astrosphere to avoid contamination of the background by the SWCX photons. We note, though, that in the tests where we used the full outer annulus as a background, we got very similar estimates of the MLRs, which indicates that SWCX photons are indeed contained in the annulus closer to the star that we used as a CX source. We also mention that the observation 0748010101 of $\epsilon$~Eri in Large Window had a non-standard background, namely, a large box placed far away from the central star. This was because there was no space in the camera's FOV for both the astrospheric annulus and the background annulus.

\section{Discussion}
\label{sec:discussion}

\begin{figure*}
\centering
\includegraphics[width=1.0\columnwidth]{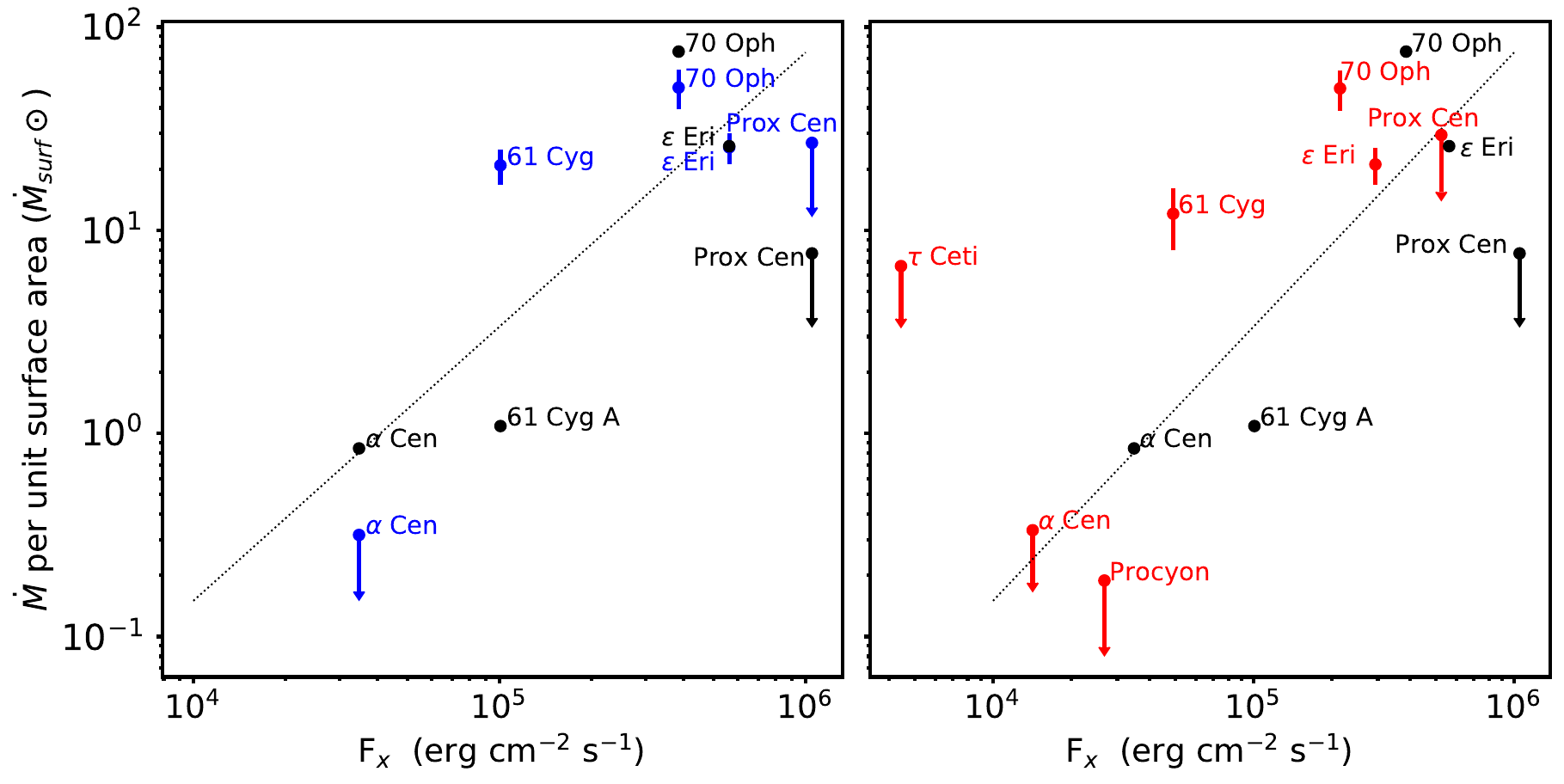}
\caption{Mass loss rates per unit surface area in solar units as a function of the surface X-ray flux. The dotted line shows a power-law fit to the detections using the hydrogen wall method\cite{Wood14} (black points). No error bars were provided for that work but systematic uncertainties are likely similar to those for our study. For our study, the data are presented as mean values $\pm 1 \sigma$. The left panel shows the MLRs per surface unit area from our study calculated using the same X-ray fluxes and stellar radii as in the hydrogen wall method\cite{Wood02,Wood05} (blue dots; $\tau$~Cet was not in the sample), while the right panel shows them calculated using the data from Table~\ref{t:results} (red dots). We do not include stars beyond the ``wind-divide'' line with $F_X > 10^6$~erg/cm$^2$/s (except Prox~Cen, for which we only obtain an upper limit) due to absence of suitable observations in the XMM-Newton archive.  }
\label{fig:MdotFx}
\end{figure*}

Our three confident (2.6--6 sigma) detections of stellar astrospheres in X-rays correspond to high mass loss rates. These MLRs are in approximate agreement with previous inferences for mass-loss rates of these stars from ISM hydrogen wall Ly-$\alpha$ line absorption\cite{Wood02,Wood05,Wood14}. As a summary of our results, Fig.~\ref{fig:MdotFx} presents the dependence of our mass loss rate estimates on the surface X-ray flux $F_X$ of our sample stars. The figure also shows a power-law fit to the data obtained using the hydrogen wall method\cite{Wood14}. One can see that our results are in a good agreement taking into account all uncertainties. We note that the hydrogen wall method also indicated a presence of a ``wind divide'' line with stars with very high surface $F_X$ fluxes showing low mass loss rates per unit area\cite{Wood14}, but this divide is beyond the limits of the $F_X$ flux shown in Fig.~\ref{fig:MdotFx} (with a possible exception of Prox~Cen, for which we only report an upper limit). We plan to investigate these stars using our method in future studies.

In addition to observations, multiple models for stellar mass loss rates based on observations of the solar wind are available. Some of them are based on observations of spin-down of large numbers of stars in stellar clusters\cite{Bouvier97,Johnstone15,Evensberget21}, stellar evolution models\cite{Cranmer19,Vidotto21}, and magnetic field topology\cite{Cohen14,Chebly23}. The spin-down rates are well constrained; however, there are still open questions regarding the effectiveness of the removal of the star's angular momentum by its stellar wind, which has direct implications for estimated mass loss rates\cite{Weber67}. Other models simulate stellar outflows based on the dissipation of Alfv\'en waves in stellar coronae\cite{Alvarado-Gomez16,Reville20}. However, variations in Alfv\'en wave energy influence the wind properties and the resulting mass loss rates very significantly\cite{Kavanagh21}. 
In addition to that, solar wind simulations carried out using the low-resolution magnetograms of the three stars instead of the actual solar high-resolution magnetogram can lead to an order of a magnitude difference in the simulated solar wind properties\cite{Jardine17,BoroSaikia20}. Direct estimates of stellar mass loss rates are crucial to benchmark these models.

We have presented here the first detections of astrospheric signals in X-rays, but we conclude that such detections are still difficult. They require high-quality spectra, low background, and very good angular resolution to resolve the faint astrospheric signals. Although it is relatively easy to conclude if there is an excess signal in the oxygen K$\alpha$ line, the interpretation of the results and the MLR estimates depend on the preliminary knowledge of the properties of the ISM surrounding the stars. Our results emphasize the importance of high-quality modeling of stellar winds and their interaction with the surrounding ISM, and of studies of the ISM itself.

\textbf{Conclusions.} Detections of X-ray emission from stellar winds are pivotal for refining the stellar wind models, models of atmospheric loss, and understanding the evolution of MS stars. Our results present the first detection of a SWCX signal from stellar astrospheres of Sun-like stars, confirming the potential of this new method, with more detections hopefully to follow with future high performance observatories such as Athena.

\section*{Methods}

Searching for astrospheric signals can be a non-trivial task. Although stellar astrospheres are large objects, X-ray emission produced there is quite faint and is dwarfed by the emission of the central source. For this reason, in our analysis we focused only on nearby stars located withing approximately 5~pc from the Sun. 

Due to characteristics of the XMM-Newton observatory, X-ray emission from an annulus around the central star is usually dominated by the stellar spectrum due to the instruments' broad wings of the point-spread-function (PSF). Although more than 90 percent of emission are concentrated within approximately 50$^{\prime\prime}$ of the central source (\url{https://xmm-tools.cosmos.esa.int/external/xmm_user_support/documentation/uhb/onaxisxraypsf.html}), in our analysis of very bright nearby stars, traces of the bright central source's emission could be detected almost everywhere in the image. To eliminate the contamination by the PSF, we have developed a procedure that allowed us to look for the \textit{excess} emission in addition to the stellar PSF emission. We have focused on searching for the excess emission in the prominent K$\alpha$ triplet at 0.56~keV. CX reactions produce a large variety of lines, but not all of them are equally bright. Oxygen is the most abundant heavy ion in the solar wind and the K$\alpha$ triplet is one of the brightest CX lines observed in solar system objects. Likely, the same assumption can be made about winds of other Sun-like stars and their astrospheres. Due to low brightness of the extended astrospheric emission and the limited spectral resolving power of the CCD detectors, the quality of the data in the extended regions does not allow us to reliably detect lines of other elements. 

\textbf{To detect X-ray photons} produced by CX in stellar astrospheres, we developed the following method:
\begin{enumerate}
\item We extracted PN, MOS1, and MOS2 spectra of the point-like stellar source following the standard procedure presented in the SAS data analysis threads. This includes the generation of the appropriate response function (rmf) and the ancillary response function (arf) containing the effective area function. 
\item We modeled the central star's emission spectrum using \texttt{Xspec} tools and the model vapec+vapec+vapec that represents a three-component plasma in collisional ionization equilibrium and with free element abundances. Examples of our fits for the central stars' spectra are shown in Fig.~\ref{fig:StarsSpectra}. 
\item We prepared the model of the stellar spectrum obtained in the previous step for comparison to the emission detected in the annuli. We froze all parameters of the 3T vapec model once they have been fitted to the stellar spectrum and only allowed the overall normalization (amplitude) of the model to vary, while the ratio between the three vapec norms was kept fixed to retain the shape of the model spectrum. 
\item We fitted the model from the previous step to the spectra of the annuli. In most cases, a scaled-down model of the stellar emission reproduced the spectra of the annuli quite well and there was no excess emission. This means that the emission in the annuli was dominated by the instrument's PSF, that is, by the emission of the central star. But in several cases, we did see excess emission around 0.5-0.7~keV, which we interpreted as an indication of excess astrospheric emission produced by CX.
\item To quantitatively estimate the amount of excess emission, we added a Gaussian line centered at 0.56~keV to our 3T vapec model in which all parameters were frozen. This energy corresponds to the energy of the oxygen K$\alpha$ triplet line. We used the standard \texttt{gauss} model available in \texttt{Xspec} for the Gaussian line. We fitted the model again, allowing the norm of the emission from the Gaussian line to vary,  apart from the normalization of the underlying stellar model, but keeping the line energy and the line width $\sigma = 0.0$ constant. 
\item We compared the fit statistics for the scaled-down model of the stellar spectrum and the same model with the additional Gaussian line. We only considered the signal as an astrospheric detection if adding the Gaussian line improved the fit for all detectors fitted together (where all three were available), and if the errors of the flux did not exceed the flux itself. We considered a decrease of the $\chi^2$ by $>$2 units as a reliable detection (see Table~\ref{t:annuli} for $\chi^2$ statistics).
\item To estimate the flux originating in the astrosphere, we next set both norms of the main model to zero, keeping only the Gaussian line. Then, we run the \texttt{flux} command in the range 0.5-0.7~keV which yielded the number of excess counts and the excess flux in erg/s/cm$^2$ that we interpreted as the astrospheric signal.
\end{enumerate}

\textbf{Influence of the stellar model.} To check if the model we used for stellar spectra influenced our results, we have also performed our analysis with two additional models: 1) 2T model; 2) c6vmekl, a differential emission measure model that uses Chebyshev representation function of T with multi-temperature mekal models. We have obtained very good agreement for mass loss rates estimated using 2T and 3T models of the stellar spectra. While the model c6vmekl could satisfactorily reproduce the emission from hot stellar coronae, it had bad $\chi^2$ for stars with cooler coronae such as Alpha Centauri and Procyon that emit in in a narrower range of T above approximately one million Kelvin. Nonetheless, for all observations for which c6vmekl could produce a satisfactory fit of the stellar spectrum, the mass loss rate estimates we obtained with this model were quite similar to the 2T and 3T models. In the final version of our estimates presented in the article, we used the 3T model as having the best $\chi^2$ characteristics for the considered stellar spectra. 

\textbf{Background choice} is an important decision that can influence the results, especially when the signals of interest are very faint. In our analysis, we have tested multiple backgrounds, for instance circular regions located far away from the central star. The smaller the background, the more influence it had on the results, because for small background areas, their spectra can be dominated by a spectrum of a random faint source unrelated to the stellar system of interest. For this reason, we decided to use large backgrounds in our analysis. The default background for most stars was the outer annulus (see Fig.~\ref{fig:StarsRegions}). For 70~Oph, $\epsilon$~Eri, $\alpha$~Cen and Prox~Cen, we have used one half of the outer annulus as the background. This is because we selected the background based on the positional angles of the astrospheres of these stars, namely, the half-annulus covering the upwind direction (Fig.~\ref{fig:StarsRegions} and Table~\ref{t:results}, Brian Wood, private communication). We have tested the spectra of full background annuli as backgrounds and obtained generally the same results as shown in Fig.~\ref{fig:Annuli}. For other stars, the position angles of the astrospheres were not available, so that we used the full far annuli as backgrounds. In general, our results were robust and did not depend on the background choice as long as the background area was comparable in size to the annuli area. Important note: since the background was always observed simultaneously with the sources of interest, there was no remnant SWCX emission from our own heliosphere that could be responsible for the excess 0.5-0.7 keV emission because such emission is diffuse and extended across the FOV. 

\textbf{Influence of the spatial distribution of CX emission on MLR estimates.} Strong stellar winds tend to sweep neutral gas to the edges of the astrosphere which leads to pile-up of neutral gas near the edges of the system. Studies of MLRs based on Lyman-$\alpha$ absorption showed that most of this absorption does originate in this hydrogen wall region\cite{Wood05}. It is predicted that X-ray CX emission should also be stronger in this region\cite{Wargelin02}. In our analysis, we have tested different widths and experimented with using several adjacent annuli with increasing radii, the results from which we co-added. For instance, we used five or four annuli instead of one. Although we arrived at similar mass loss rates, the errors in every annulus were large due to the faintness of the signal, which made the analysis less robust. For this reason, we have opted for analysing the strongest possible SWCX signal from just one very wide annulus around the star, which reduced the errors. Therefore, due to the faintness of the signal, we could not detect the presence and location of the hydrogen wall region in the images. It is possible that future missions with better spectral and spacial resolutions will provide observations for which such analysis will be possible. 

\textbf{Heliospheric and magnetospheric contaminations.} For the geocoronal, or magnetospheric emission, we have checked the orbits of XMM during each observation to see if the line-of-sight bisected the dayside region of the magnetosheath. During perigee passage around the Earth, XMM has the filter wheel in the closed position and is therefore unable to make astronomical observations. The dayside, or subsolar regions, is where one can expect higher geocoronal solar wind charge exchange emissions. Even though any resulting foreground emission, expected to be diffuse, should be removed by the background subtraction method, we have shown that even for those observations that occur on the dayside of the magnetosheath, most lines of sight will miss the areas of highest X-ray emissivity from SWCX. In addition to that, the argument originally presented for \textit{Chandra}\cite{Wargelin14} (footnote 11 of the reference) stating that within the FOV of \textit{Chandra} or XMM SWCX emissivity should not vary significantly and thus should be removed by background subtraction.

\textbf{The most spatially variable feature}, albeit over large regions, of the heliospheric emission is the He focusing cone. The He focusing cone is a region of increased density formed by the gravitational pull of the Sun on He atoms, centered around the axis of the outgoing He flow direction ($\sim 75^\circ$, $-5^\circ$ in ecliptic coordinates). The Earth crosses the cone every year in December.  The maximum emissivity region for H is farther away from the observer, in the opposite direction of the He cone towards the incoming ISM H flow. Following a careful inspection of the observation geometries of our data sample (view directions projected on the ecliptic plane overplotted in the figure), most of them are at high southern ecliptic latitudes (with the exception of 70 Oph which is at +25$^\circ$ ecliptic latitude), and none are closer than 25$^\circ$ of the He cone axis, the closest being the 61 Cyg target. Therefore, we conclude that neither the heliospheric nor magnetospheric CX signals were a likely source for contamination of the annuli spectra. Again, any expected CX signals would be spatially more extended than our annuli and threrefore be subtracted by our background removal method.

\textbf{To avoid contamination} of our results by the Out-Of-Time events appearing on the detector as a stripe during the CCD readout, we have removed them following the SAS thread dedicated to this problem (\url{https://www.cosmos.esa.int/web/xmm-newton/sas-thread-epic-oot}). We have also tested if the edges of separate CCD detectors visible on PN images (Fig.~\ref{fig:StarsRegions})  influenced the results. 
We have created spectra including and excluding these regions, and concluded that they did not influence our results. For MOS1 and MOS2 detectors, the edges of the CCD detectors were not brighter than the rest of the CCDs outside of the bright stellar PSF.

\textbf{Influence of the energy dependence of the PSFs.} Generally, point spread functions for energies in a small energy range of 0.4--1~keV were similar for all detectors. We have compared the PSFs for 0.4, 0.6, and 1.0~keV for all three detectors for the observations yielding SWCX detections using the \texttt{psfgen} routine, the encircled energy values available in the \texttt{ccf} files, and \texttt{calview}. For the PN detector, the largest PSF was for the smallest energy. Therefore, if the excess emission was originating from the PSF, the largest excess would be observed near 0.4~keV energy and the shape would likely not resemble a Gaussian line with the maximum emission near 0.56~keV. For MOS1 and MOS2, PSFs were very similar for these three energies with the differences not significant enough to produce observed excess emission.

\textbf{The next important step} was converting the astrospheric fluxes in the K$\alpha$ line to stellar mass loss rates. Since excess emission is produced by CX between heavy ions of the stellar wind and the neutral ISM (strictly speaking: neutrals penetrating the astrosphere from the ISM), understanding the ISM surrounding the stars was crucial for estimating stellar mass loss rates. Table~\ref{t:annuli} summarizes the annuli size and $\chi^2$ statistics for each observation excluding ($\chi^2$) and including ($\chi^2_{0.56~keV}$) the additional Gaussian line. The difference is highly significant for 70 Oph and for the first (Full Frame) observation of $\epsilon$~Eri. 

We have also estimated the mass loss rates based roughly on the sizes and densities of reconstructed astrospheres in the hydrogen wall methods, and obtained similar mass loss rates, although they were in general somewhat larger than the ones presented in Table~\ref{t:results}. This discrepancy can be explained by uncertainties in the parameters in Eqs.\ref{e:wind} and \ref{e:mdot}.

\textbf{We have included} the following XMM-Newton observations in our analysis: 70~Oph (observation IDs 0044740101, 0044741301, 0044741401), $\epsilon$~Eri (0112880501 -- Full Frame, 0748010101 -- Large Window), $\alpha$~Cen (0550060901, 0550061001, 0670320301, 0690800301, 0741700301, 0760290301), Prox~Cen (0551120201, 0551120301, 0801880501), Procyon (0415580101, 0415580201, 0415580301), 61~Cyg (0690800901, 0741700601, 0801871001), $\tau$~Cet (0670380501). Several additional observations were available that we excluded from our analysis for various reasons: the observations 0801870301, 0823500301 of $\alpha$~Cen  (due to erroneous generation of response matrixes for MOS2 that made reliable fitting of all three cameras together impossible), 0783340301 (due to flaring particle background), and one observation 0202610401 of 61~Cyg (due to problems producing event lists).

We have extracted the spectra of annuli located at the same distances from the host star in all observations of a particular star, with the exception of $\epsilon$~Eri where the Large Window observation had a smaller FOV. Most observations were of high quality and had very low background levels. We have followed the SAS examples for the cutoff thresholds of rates less than 0.35 counts/second and 0.4 counts/second for EPIC-MOS and EPIC-PN, respectively. The only observation that has been heavily affected by the background noise to the extent that we were not able to obtain any annulus spectra was the observation 0783340301 of $\alpha$~Cen. 

For several observations, some areas in the FOV of the MOS1 and/or MOS2 cameras were missing, even though the central source was always available for the analysis. In these cases, we could not obtain spectra of annuli for MOS1 and MOS2. This was a case for following observations: all Procyon observations, $\alpha$~Cen obs. ID 0741700301 (MOS2 observations affected),  $\alpha$~Cen obs. ID 0760290301 (only a fraction of the annulus was present in MOS1 and MOS2 FOVs), and all Prox~Cen observations. For $\alpha$~Cen observations 0760290301 and 0741700301 it was still possible to fit the spectra of PN, MOS1, and MOS2 cameras together despite gaps in FOVs. For other cases, we have relied on PN observations to obtain mass loss rates, because extracted spectra of MOS1 and MOS2 did not contain enough counts for the spectral analysis. All 70~Oph observations and the observation of $\epsilon$~Eri in the Full Frame mode had images from all three detectors available. In 61~Cyg observations, PN and MOS2 were fully available, and MOS1 had small gaps in the FOV. The observation of $\epsilon$~Eri in Large Window mode had large gaps in MOS1 and MOS2 images, so we have relied on PN observations to estimate the mass loss rate. 

For 70~Oph, $\epsilon$~Eri (Full Frame), $\alpha$~Cen and 61~Cyg, we have used spectra of PN, MOS1, and MOS2 detectors fitted together to estimate mass loss rates to obtain the numbers shown in Table~\ref{t:results}. For $\alpha$~Cen, we used other observations where only the PN detector was available to verify our results, and got similar upper limits for mass loss rates, although they had a larger scatter. For $\epsilon$~Eri (Large Window), Procyon, Prox~Cen, and $\tau$~Cet, we had to rely on PN detector only. 

\textbf{To illustrate our method}, we list below the commands we used to generate the spectra of the observation of 70~Oph ID 0044740101 shown in Figs.~\ref{fig:StarsRegions}, \ref{fig:StarsSpectra}, and \ref{fig:Annuli}. We have used the standard list of commands to create event lists cleaned and filtered for particle background for PN, MOS1, and MOS2, and to extract the spectrum of the OoT events. Since OoT events were only visible in PN images, we have not removed them from the MOS1 and MOS2 spectra. In the list of commands below, the files \texttt{EPICclean\_EPN.fits}, \texttt{EPICclean\_MOS1.fits}, and \texttt{EPICclean\_MOS2.fits} are the event files cleaned and filtered for particle background, and \texttt{P0044740101PNS002OOEVLI0000.FIT} is the OoT event list.

First, we have selected the spectra of the central star (source), the annulus and the background. We used DS9 physical coordinates to select the regions. Note that the background is actually a half-annulus, as illustrated in Fig.~\ref{fig:StarsRegions}:

\begin{enumerate}

\item \texttt{evselect table=EPICclean\_EPN.fits withspectrumset=yes \\ 
spectrumset=PN\_source\_spectrum.fits energycolumn=PI spectralbinsize=5 \\
withspecranges=yes specchannelmin=0 specchannelmax=20479 \\
    expression='(FLAG==0) \&\& (PATTERN<=4) \\ \&\& ((X,Y) IN circle(26916.772,27572.025,627.08))'}

\item \texttt{evselect table=EPICclean\_EPN.fits withspectrumset=yes \\ spectrumset=PN\_annulus\_spectrum.fits energycolumn=PI spectralbinsize=5 \\
 withspecranges=yes specchannelmin=0 specchannelmax=20479  \\
    expression='(FLAG==0) \&\& (PATTERN<=4) \\ \&\& ((X,Y) IN annulus(26916.772,27572.025,1200,10500))'}

\item  \texttt{evselect table=EPICclean\_EPN.fits withspectrumset=yes \\ spectrumset=PN\_background\_spectrum.fits energycolumn=PI spectralbinsize=5 \\ 
    withspecranges=yes specchannelmin=0 specchannelmax=20479 \\ 
    expression='(FLAG==0) \&\& (PATTERN<=4) \\ \&\& ((X,Y) IN annulus(26916.772,27572.025,10600,13855) \&\& (X,Y) IN box(27033.545,20395.943,28484.145,13678.92,4.1999757))' }

\end{enumerate}

This set of commands creates the files \\
\texttt{PN\_source\_spectrum.fits}, \\
\texttt{PN\_annulus\_spectrum.fits}, \\
\texttt{PN\_background\_spectrum.fits} \\

Then, we have made the same selection for OoT events for the source region, annulus, and the background. We show the command for the source region as an example:

\texttt{evselect table=P0044740101PNS002OOEVLI0000.FIT withspectrumset=yes \\ spectrumset=PN\_oot\_source\_spectrum.fits energycolumn=PI spectralbinsize=5 \\
 withspecranges=yes specchannelmin=0 specchannelmax=20479 \\
    expression='(FLAG==0) \&\& (PATTERN<=4) \\ \&\& ((X,Y) IN circle(26916.772,27572.025,627.08))'}

This set of commands creates the files \texttt{PN\_oot\_source\_spectrum.fits}, \\
\texttt{PN\_oot\_annulus\_spectrum.fits}, \\
\texttt{PN\_oot\_background\_spectrum.fits} \\

Then, we have subtracted the spectrum of OoT events from the spectra. Again, we show the set of commands for the source region. The commands for each annulus are analagous:

\texttt{fparkey value=CTS\_OOT fitsfile=PN\_oot\_source\_spectrum.fits+1 keyword=TTYPE2}

\texttt{faddcol infile=PN\_source\_spectrum.fits+1 colfile=PN\_oot\_source\_spectrum.fits+1 colname=CTS\_OOT     }

\texttt{fcalc clobber=yes infile=PN\_source\_spectrum.fits+1 
\\ outfile=PN\_source\_spectrum.fits clname=CTS\_OOT expr=CTS\_OOT*0.063 } 
        
\texttt{fcalc clobber=yes infile=PN\_source\_spectrum.fits+1 
\\ outfile=PN\_source\_spectrum.fits clname=COUNTS expr=COUNTS-CTS\_OOT}

Now we have the same files as above, but with OoTs removed. We ran the set of \texttt{backscale} commands for the source, annuli and background spectra (the commands for the annulus and the background look analagous to the one shown below):

\texttt{backscale spectrumset=PN\_source\_spectrum.fits badpixlocation=EPICclean\_EPN.fits}

Next, we generated rmf response and arf ancillary response files for the source spectrum:

\texttt{rmfgen spectrumset=PN\_source\_spectrum.fits rmfset=PN.rmf}

\texttt{arfgen spectrumset=PN\_source\_spectrum.fits arfset=PN.arf withrmfset=yes rmfset=PN.rmf 
    badpixlocation=EPICclean\_EPN.fits detmaptype=psf}
    
Note that we have used the flag \texttt{detmaptype=psf}. This set of commands generates \texttt{PN.rmf} and \texttt{PN.arf}. For the annulus, we have generated a separate ancillary response file \texttt{PN\_annulus.arf}:

\texttt{arfgen spectrumset=PN\_annulus\_spectrum.fits arfset=PN\_annulus.arf \\
withrmfset=yes rmfset=PN.rmf badpixlocation=EPICclean\_EPN.fits \\
extendedsource=yes detmaptype=flat}
    
Note that we have used different flags \texttt{extendedsource=yes} and \texttt{detmaptype=flat} to create arf files for the extended source. 
Finally, one can rebin (``group'') the spectra for appropriate statistical handling in \texttt{Xspec}:

\texttt{specgroup spectrumset=PN\_source\_spectrum.fits mincounts=25 oversample=3 rmfset=PN.rmf \
    arfset=PN.arf backgndset=PN\_background\_spectrum.fits 
    \\ groupedset=PN\_source\_spectrum\_grp.fits }

\texttt{specgroup spectrumset=PN\_annulus\_spectrum.fits mincounts=25 oversample=3 rmfset=PN.rmf \
    arfset=PN\_annulus.arf backgndset=PN\_background\_spectrum.fits \\ groupedset=PN\_annulus\_spectrum\_grp.fits }

This set of commands creates the spectra \texttt{PN\_source\_spectrum\_grp.fits} and\\ \texttt{PN\_annulus\_spectrum\_grp.fits}. The file \texttt{PN\_background\_spectrum.fits} is used as a background. These spectra can be fitted and processed using standard \texttt{Xspec} routines.

The spectra for MOS1 and MOS2 cameras can be created in a similar way, but one should use different flags for \texttt{evselect} command (an example for MOS1 source spectrum), and the OOT events do not have to be removed:

\texttt{evselect table=EPICclean\_MOS1.fits withspectrumset=yes \\
spectrumset=MOS1\_source\_spectrum.fits energycolumn=PI spectralbinsize=5 \\
withspecranges=yes specchannelmin=0 specchannelmax=11999  \\ 
   expression='\#XMMEA\_EM \&\& (PATTERN<=12)\\ \&\& ((X,Y) IN circle(26916.772,27572.025,627.08))'}

\section*{Data availability}

All data used for this study is publicly available in XMM-Newton data archive at \url{https://nxsa.esac.esa.int/nxsa-web/}.

\section*{Code availability}

We have used the standard tools developed for the data reduction and calibration of XMM-Newton observations, the XMM-Newton Science Analysis System (SAS) (\url{https://www.cosmos.esa.int/web/xmm-newton/sas}). We have used the version \texttt{xmmsas\_20201028\_0905-19.0.0} of the SAS and followed standard procedures for extraction of spectra of point-like sources as described in \url{https://www.cosmos.esa.int/web/xmm-newton/sas-threads}. For science analysis, we applied the X-Ray Spectral Fitting Package (\texttt{Xspec}) (\url{https://heasarc.gsfc.nasa.gov/xanadu/xspec/}) created by the NASA's High Energy Astropyhsics Science Archive Research Center (\url{https://heasarc.gsfc.nasa.gov/}). We have used the publicly available Python library \texttt{Matplotlib} for plotting.

\section*{Acknowledgements}
K.K. and M.G. acknowledge the support by the Austrian Research Promotion Agency (FFG) Project 873671 ``SmileEarth''. D.K. acknowledges the support by the CNES. J.A.C is supported by Royal Society grant DHF\textbackslash R1\textbackslash 211068. The authors are grateful to Dr. Brian E. Wood for calculating position angles of several astrospheres. 
\vspace{1em}

\section*{Author Contributions}
\small
K.G.K.\ hatched the original idea of the manuscript and performed the majority of data reduction and analysis. 
M.G.\ contributed equally with K.G.K. to data interpretation.
D.K.\ contributed with expertise of SWCX emission analysis and modeling in the heliosphere and by extension to astropheres.
J.A.C.\ provided their expertise for analysis of extended sources and instrumental effects.
C.L.\ provided the final insight into the data's interpretation that helped to refine the mass loss estimates and put them in context.
S.B.S.\ contributed her expertise on current state-of-the-art of stellar wind modeling and observations.
All authors contributed to the text.

\newpage

\clearpage
\newpage
\bibliographystyle{naturemag} 

\end{document}